\documentclass{iaus}
\usepackage{graphicx}

\title[Globular cluster black hole] 
{An X-ray emitting black hole in a globular cluster}

\author[Maccarone et al.]   
{T.J. Maccarone$^1$,G. Bergond$^2$,A. Kundu$^3$, K.L. Rhode $^{4,5,6}$, J.J. Salzer$^{4,5}$, I.C. Shih$^3$, S.E. Zepf$^3$}

\affiliation{$^1$School of Physics and Astronomy, University of Southampton, Southampton, UK, SO16 4ES\break email: tjm@phys.soton.ac.uk\\[\affilskip]
$^2$Instituto de Astrof\'\i sica de Andaluc\'\i a (IAA/CSIC), Camino Bajo
de Hu\'e tor 50, 18008 Granada, Spain\\[\affilskip]
$^3$ Department of Physics and Astronomy, Michigan State University, East Lansing, MI 48824, USA\\[\affilskip]
$^4$ Department of Astronomy, Indiana University, Bloomington, IN 47405, USA\\[\affilskip]
$^5$ Department of Astronomy, Wesleyan University, Middletown, CT, 06459, USA\\[\affilskip]
$^6$ Department of Astronomy, Yale University, New Haven CT, 06520, USA\\[\affilskip]}

\pubyear{2007}
\volume{246}  
\pagerange{}
\date{}
\jname{Dynamical Evolution of Dense Stellar Systems}
\editors{E. Vesperini, M. Giersz, A. Sills}
\begin{document}

\maketitle

\begin{abstract}
We present optical and X-ray data for the first object showing strong
evidence for being a black hole in a globular cluster.  We show the
initial X-ray light curve and X-ray spectrum which led to the
discovery that this is an extremely bright, highly variable source,
and thus must be a black hole.  We present the optical spectrum which
unambiguously identifies the optical counterpart as a globular
cluster, and which shows a strong, broad [O III] emission line, most
likely coming from an outflow driven by the accreting source.

\keywords{}
\end{abstract}

\firstsection 
\section{Introduction}

Since the early days of X-ray astronomy, there has been considerable
debate over whether globular clusters contained black holes.  With the
discovery of Type I X-ray bursts from all globular clusters in the
Milky Way with bright X-ray sources (starting with Grindlay et
al. 1976), and their subsequent explanation as episodes of
thermonuclear burning on the surfaces of neutron stars (Woosley \&
Taam 1976; Swank et al. 1977), it became clear that there was no
evidence for any accreting black holes in the Milky Way's globular
cluster system.

Intepretations of the observations have been taken in two directions.
One is simply that given only 13 bright X-ray sources in the Milky
Way's globular cluster system, it is not so unlikely for them all to
have neutron star accretors, especially in light of the fact that
about 10 times as many neutron stars as black holes are expected to be
produced for most stellar initial mass functions.  The alternative is
that dynamical effects are responsible for ejecting black holes from
globular clusters.  Severe mass segregation is likely to take place
for globular cluster black holes, as they should be many times heavier
than all the other stars in the cluster.  This can lead to the
formation of a ``cluster within a cluster'' where the heaviest stars
(i.e. the black holes) feel negligble effects from the other stars in
the cluster, which in turn leads to a cluster with a short evaporation
timescale (Spitzer 1969).  Numerical calculations have found that this
evaporation can be accelerated further due to binary processes (e.g
Portegies Zwart \& McMillan 2000).

Early results from the Chandra X-ray Observatory gave new hope that
globular cluster black holes might be detectable, by opening up the
window of looking in other galaxies.  Previously, only ROSAT could
resolve point sources in other galaxies, and its localization of
sources was generally not good enough to allow for unique
identification of optical counterparts.  The first few years of
Chandra observations revealed several extragalactic globular cluster
X-ray sources brighter than the Eddington limit for a neutron star
(e.g. Angelini et al. 2001; Kundu et al. 2002), but a globular cluster
may contain multiple bright neutron stars (as does, for example M~15
in our own galaxy -- White \& Angelini 2001), and that the quality of
X-ray spectra available from Chandra for even the brightest
extragalactic sources is insufficient to make phenomenological
determinations that a source has a black hole accretor.  It was
pointed out that only large amplitude variability could prove that we
were seeing the emission from a single source, rather than multiple
sources (Kalogera, King \& Rasio 2003).

Furthermore, the optical catalogs used to identify globular cluster
counterparts to X-ray sources have been predominantly photometric
catalogs, sometimes made even without color selections being used to
ensure that that the optical source in question truly is a globular
cluster.  Most studies done to date have focused on HST images of the
central regions of elliptical galaxies with high specific frequencies
of globular clusters.  In these regions, and with the angular
resolution of HST, contamination will be rare, especially if color
cuts are used to ensure that the contribution of background quasars is
minimized.  In the halos of galaxies, the surface density of real
globular clusters will drop, and contamination will be a more serious
problem.  In either case, when one is looking for conclusive proof
that an object is a globular cluster black hole, spectroscopic
confirmation that the object is a globular cluster is essential -- the
fractional contamination of the X-ray sources due to background AGN
will be more serious at very high fluxes, corresponding to
luminosities above $10^{39}$ ergs/sec than it will at lower levels
consistent bright neutron star accretors.  Furthermore, the optical to
X-ray ratios for globular cluster black holes near the Eddington limit
and background quasars are quite similar.

\section{Discovery of a globular cluster black hole}

In a recent paper, we found a source meeting all the strict criteria
for identifying an X-ray emitting black hole in a globular cluster
(Maccarone et al. 2007).  XMMU J1229397+075333 has an X-ray luminosity
of 4.5 $\times10^{39}$ ergs/sec.  It shows variability of a factor of
7 in count rate, in a time span of about 3 hours (see Figure 1).  It
is located in a spectroscopically selected globular cluster.  The
object was first detected by ROSAT, and is included in the
intermediate luminosity X-ray source catalog of Colbert \& Ptak
(2002), but as it is about 7 arcminutes from the center of NGC~4472,
its host galaxy, good optical follow-up, combined with the positional
accuracy obtained from archival Chandra data were necessary to ensure
that it was not just a background active galactic nucleus.

\begin{figure}
\center{\includegraphics[width=5.8cm]{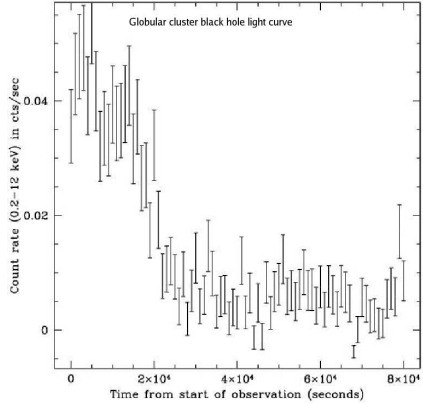}}
\caption{The XMM-Newton light curve of XMMU J1229397+07533.}
\end{figure}

\section{Observational properties}

This source has several unusual X-ray properties, all of which give
some clues about its possible nature.  Its X-ray spectrum is dominated
by a very soft, quasi-thermal component.  The best fitting models give
inner disk radii of thousands of kilometers, with temperatures of
about 0.2 keV.  The change in X-ray count rate is predominately a
change in the low energy X-ray emission, and is consistent with a
change only in foreground absorption.  This source was also observed
in 1994 by ROSAT, and in 2000 by Chandra, in both cases as part of
observations of the whole of NGC~4472.  In both of these cases, the
X-ray luminosity was within a factor of a few of $4\times10^{39}$
ergs/sec, with bigger uncertainty in the luminosities than for XMM due
to the lower count rates, narrower spectral energy range, and, in the
case of ROSAT, much poorer spectral resolution, when compared with
XMM.  This suggests strongly that this source has been persistently
bright for at least a 12 year period.  The comparison of the different
epochs spectra and detailed spectral fitting can be found in Shih et
al. (2007).

The optical spectrum also shows an unusual feature which may shed
light on the nature of the accretor.  Two optical spectra have been
taken of this object, and in each case, a strong, broad [O III]
emission line is seen (see figure 2 for one example).  Unfortunately,
the observations were aimed at making spectroscopic confirmation of
the globular cluster nature of a large number of clusters, and at
doing kinematic studies of the clusters.  This means that the spectra
have been taken with relatively high spectral resolution, but over a
relatively narrow wavelength range, and this wavelength range includes
only [O III]5007 \AA among the emission lines which are commonly
strong.  Emission lines are rarely seen in globular clusters'
integrated spectra, and are normally attributed to planetary nebulae
when they are seen.  The line is many times broader in velocity
profile than lines from planetary nebulae.  A detailed description of
the line properties can be found in Zepf et al. (2007).

\begin{figure}
\center{\includegraphics[width=5.5 cm]{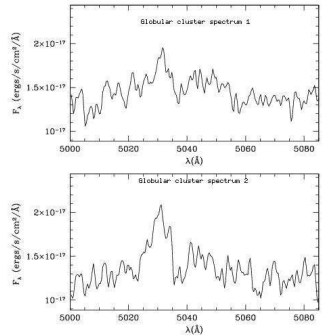}}
\caption{The optical spectrum of RZ2109, the globular cluster
containing XMMU J1229397+07533.  The emission line appears slightly
redshifted from the position of [O III] at 5007 \AA, with the redshift
the same for the line as it is for the absorption lines in the stellar
continuum.  The two different epochs of VLT spectra are shown.}
\end{figure}

\section{Interpretation: the nature of this object}

The large inferred inner disk radius from the X-ray spectrum of this
object implies that one of two things is most likely to be true.
Either the object contains an intermediate mass black hole of a few
hundred solar masses accreting at about a few percent of its Eddington
rate, or it containts a stellar mass black hole accreting at a mildly
super-Eddington rate.  In the former case, the combination of observed
temperature and luminosity is straightforward, as it requires only
that the accretion disk extend in to the innermost stable circular
orbit for the black hole.  

The latter case, while not as intuitively obvious, is equally
plausible physically.  In the event of super-Eddington accretion, one
can expect annuli where the luminosity is the local Eddington
luminosity.  This yields a total luminosity of the Eddington
luminosity multiplied by the logarithm of the ratio of the mass
accretion rate to the accretion rate needed to reach the Eddington
luminosity (see e.g. Begelman, King \& Pringle 2006).  While the
spectrum will not any more be exactly that predicted by the standard
disk blackbody model, with moderate quality data like those which
exist for extragalactic X-ray binaries, this model will provide an
acceptable fit, and will yield an inner disk radius equal to the
radius of the outermost annulus where the source is locally Eddington
limited.  Taking this into account, our data are well fit by a model
in which the black hole mass is about 10 $M_\odot$, and the accretion
rate is about 40 times what would be needed to produce an
Eddington-luminosity source (see e.g. Begelman, King \& Pringle 2006).

The variability can be explained by a puffed up warped region in the
accretion disk which precesses, sometimes obscurring the central
region of the accretion flow, while, most of the time, it is not
blocking our line of sight to the source (see Shih et al. 2007 for a
more detailed discussion of the source's X-ray variability).  A
precession period of about 100 days, combined with a radius of about
$10^{11}$ cm for the location of the warp, gives a good match to the
timescale of about 3 hours on which the obscurration takes place. One
could alternatively consider self-obscurration by a variable disk
wind, but it seems unlikely that the observed column density would
change from a level which is negligible compared to the foreground
column density of 1.6$\times10^{20}$ cm$^{-2}$ to the level of
approximately $3\times10^{21}$ needed to match the spectrum in the
faint part of the light curve.  Strongly warped accretion disks which
are puffed up enough to allow self-obscurration are most likely to be
seen for luminosities near the Eddington luminosity, so the
interpretation of the source as a super-Eddington stellar mass black
hole is favored in this scenario.  Given the long inferred precession
timescale, despite the high mass accretion rate which speeds the
transfer of warp angular momentum, we expect such a system to have a
relatively long orbital period of $\sim$1 month.  This is in good
agreement with the finding that the object has been a bright X-ray
source for over a decade, as long period systems show long outbursts
followed by extremely long periods of quiescence (e.g. Portegies
Zwart, Dewi \& Maccarone 2005 and references within).
 
The optical spectrum is consistent with the idea that this system
contains a super-Eddington stellar mass black hole, rather than an
intermediate mass black hole (see Zepf et al. 2007 for a more detailed
discussion of the cluster's optical spectrum).  The broad [O III]
emission line can be explained as the result of a bubble being
inflated by strong disk winds from a stellar mass black hole exceeding
the Eddington limit.  This bubble then collides with the interstellar
medium in the globular cluster, producing a shock.  The shock velocity
will approximately equal the physical width of the emission line.  The
shock velocity can be slowed down from the original outflow velocity
to the observed velocity in a reasonable duration, given expected
density of the interstellar medium in a globular cluster.  Other
scenarios for producing a broad line, especially from an intermediate
mass black hole, run into problems -- while [O III] can come from
photoionization, and the line width could be due to virial motions
near the black hole, it is difficult (albeit not impossible) to allow
for enough material to be located close enough to the black hole to
match both the line luminosity and the line width.  Thus, an outflow
would still be required in the case of an intermediate mass black
hole, and outflows are generally quite weak for sources with
luminosities well below their Eddington limits (Proga 2007).

\section{Conclusions}
We have shown for the first time clear evidence of a globular cluster
black hole, on the basis of strong, highly variable X-ray emission
from a source in a spectroscopically confirmed globular cluster.
Based on the X-ray spectrum, the characteristic variability, and the
[O III] emission in the optical spectrum, the black hole is most likely
a stellar mass object accreting far faster than its Eddington rate.
Given that it is difficult to develop a scenario in which a globular
cluster could have both a stellar mass black hole in a binary and an
intermediate mass black hole, this argues against the idea that all
globular clusters contain intermediate mass black holes.  This
discovery also motivates future searches for quiescent stellar mass
black holes in the Milky Way's globular clusters; these may be hiding
among the X-ray sources currently classified as cataclysmic variable
stars.  Radio emission should be detectable only from quiescent
stellar mass black holes, and should be the one feasible discriminant
between the two classes of systems.

\begin{acknowledgments}
We are grateful to Tom Dwelly, Sebastian Jester, Elmar K\"ording, Phil
Charles, Mike Eracleous, Steinn Sigurdsson, Robin Barnard, Andrew
King, Guillaume Dubus and Mark Voit for useful discussions.
\end{acknowledgments}

\end{document}